# Influence of Defects on Photoconductivity and Photocatalytic Activity of Nitrogen-Doped Titania

A. A. Minnekhanov[1,2,*], N. T. Le[1], E. A. Konstantinova[1,2,3], P. K. Kashkarov[1,2,3]

[1]Faculty of Physics, M. V. Lomonosov Moscow State University, 119991 Moscow, Russian Federation

[2]National Research Center Kurchatov Institute, 123182 Moscow, Russian Federation

[3]Faculty of Nano-, Bio-, Information and Cognitive Technologies, Moscow Institute of Physics and Technology, Dolgoprudny, 141700 Moscow Region, Russian Federation

* A. A. Minnekhanov, minnekhanov@physics.msu.ru

**Abstract**

Samples of nitrogen-doped titanium dioxide (anatase, $0.2 \leq N \leq 1.0$ wt%) prepared by the sol-gel method were investigated using X-band EPR spectroscopy, photoconductivity and photocatalysis measurements. N• and NO• paramagnetic defects in $N-TiO_2$ have been observed; their concentrations and spin-Hamiltonian parameters were calculated. An increase both in the rate of the generation of free charge carriers and in the rate of photocatalysis was found in $N-TiO_2$ in contrast with non-doped titania under visible light. The correlation of the density of the observed radicals with the photoconductivity and photocatalysis data is discussed.

## 1 Introduction

Titanium dioxide $TiO_2$, known first as a white pigment widely used in industry, invariably draws researchers' attention due to its unique physical and chemical properties [1–4]. One of the most promising application areas of $TiO_2$ is photocatalysis — chemical decomposition of organic impurities, including bacteria and viruses, under illumination [5–7]. However, until recently there was a restriction of the use of $TiO_2$ as a photocatalyst, associated with the band gap width of this material (~3.2 eV for anatase): in order to achieve an acceptable rate of photocatalysis ultraviolet radiation (UV) is required. Various attempts were made to remedy this drawback: for example, surface sensitization was used where electron transfer occurs from an adsorbed dye molecule excited by visible light [1, 8, 9]. However it is doping of $TiO_2$ that has been employed the most. The rapid growth of research in this area occurred after the publication by Asahi et al. in 1999 [10], which demonstrated that nitrogen doping of $TiO_2$ leads to an effective narrowing of the band gap to the value corresponding to the visible light photon energy, due to the overlap of the 2p states of oxygen and nitrogen (it was also suggested [11], that the absorption of visible light was due to the emergence of nitrogen energy levels near the valence band edge of $TiO_2$). Subsequently, it was established that the desired result can be achieved by doping titania not only with nitrogen, but also with many other metal and non-metal elements (fluorine, carbon, etc.) [12–22]. We should also mention a similar approach to improve the photocatalytic properties of $TiO_2$, so-called self-doping, which is aimed at formation of $Ti^{3+}$ species without using any impurities [23–26]. The methods of self-doping, however, often imply high temperatures and may cause an undesirable change of titania polymorph from anatase to rutile, which is less active in photocatalytic terms.

Nevertheless, despite the large number of publications in this area doped titania capable of competing in photocatalytic efficiency with the modern purification systems has not yet been obtained. The problem lies both in the complexity of the doping procedures, which significantly raise the cost of the final product, and in the aspects of embedding impurities into the $TiO_2$ crystal lattice: there often occurs an increase not only in the rate of photogeneration of electrons and holes required for photocatalysis, but also in the rate of their recombination due to scattering off the defects of the resulting crystal structure [27–30]. In this regard, it is necessary not only to investigate in detail the electronic transport properties of new materials based on $TiO_2$, but also to determine the possible influence of the defects on them.

The aim of this work is to study the influence of defects on the photoconductivity and photocatalytic activity of nitrogen-doped nanocrystalline titania (anatase) produced by the sol-gel method. Since the most active defects in photocatalysis reactions are usually paramagnetic, electron paramagnetic resonance (EPR) spectroscopy has been chosen as the main method of our research. As complementary methods the photoconductivity and photocatalytic activity of the samples have been measured.

## 2 Experimental

Nitrogen-doped $TiO_2$ powders were synthesized by the sol-gel method using $(NH_4)_2CO_3$ as a nitrogen dopant precursor. To perform the doping reaction the precursor was boiled for 3 hours with titanic acid ($TiO_2 \cdot xH_2O$) using a reflux condenser. To obtain a series of samples with different concentrations of nitrogen impurities the following ratios of reagents were used: 20 g of ammonium carbonate in 100 ml of water, 35 g of ammonium carbonate in 100 ml of water, 75 g of ammonium carbonate in 100 ml of water. The resulting white precipitate was separated by centrifuging and annealed in air at 300° C for one day. The obtained samples have the following designations: N-$TiO_2$-1, N-$TiO_2$-2, N-$TiO_2$-3 respectively. The non-doped $TiO_2$ sample, obtained by the same method, is denoted as $TiO_2$.

To measure the photoconductivity thin films of the obtained N-TiO$_2$ samples were deposited on a quartz substrate several microns thick. The gold contacts (in the planar configuration) were deposited on the upper surface of the films using a "VUP-5" vacuum universal post. The system containing a light source (an incandescent lamp), a lattice monochromator, and lenses was used to illuminate the sample included in the electrical circuit. The produced photocurrent was registered by an HP 4192-A impedance analyzer (the frequency range was 5–10000 Hz). The sample was placed in a nitrogen cryostat, which allows measurements both at the atmospheric pressure and in the vacuum of ~10$^{-4}$ Torr. All the measurements were performed using a program written in the LabView software environment. The value of the photoconductivity $\sigma_{ph}$ was calculated as the difference between the conductivity under illumination $\sigma_{ill}$ and the conductivity in the dark $\sigma_d$: $\sigma_{ph} = \sigma_{ill} - \sigma_d$. It was assumed that the current flew uniformly over the cross-sectional area of the sample since the distance between the contacts was significantly greater than the thickness of the TiO$_2$ layer. Thus the conductivity was calculated as $\sigma = \frac{l}{ad} \Sigma$, where $\Sigma$ is the sample conductance, $l$ is the distance between the contacts, $a$ is the length of the contacts, $d$ is the sample thickness.

The EPR spectra were measured using an X-band Bruker ELEXSYS-500 spectrometer (with the working frequency of 9.5 GHz and sensitivity of 5·10$^{10}$ spin/G) in the temperature range of 77–300 K. The samples under study were placed in capillary tubes maintained in an optical cavity for the measurements in the dark or under illumination by an incandescent lamp. To calculate the concentration of paramagnetic centers (defects), the CuCl$_2$·(2H$_2$O) reference was used. The spin-Hamiltonian parameters: the hyperfine splitting, the line widths, g- and A-tensors were calculated by a computer simulation of the experimental EPR spectra using the EasySpin MATLAB toolbox [31].

An X-ray photoelectron spectroscope (XPS; Kratos Axis ultra DLD) was employed to establish the chemical composition of the samples by determining the binding energy with respect to Ti and O. The monochromatized Al$_{K\alpha}$ radiation (*hv*=1486.6 eV) was used during the experiment. The spectra were recorded with the parameter Pass Energy set at 80 using a neutralizer, compensating charging of the sample. The background correction was performed using the method, introduced by Shirley [32]. The accuracy of the measurements of the binding energy was ±0.1 eV.

The crystallite size and crystal structure of the samples were investigated using a transition electron microscope (TEM; LEO 912 AB OMEGA) with the field emission gun at 100 kV, image resolution of 0.2 nm, zoom up to 500000 times. The phase composition was examined by X-ray powder diffraction (XRD) with a DRON-3M instrument (wavelength $\lambda=1.54059$ Å (Cu K$\alpha_1$ radiation)).

The photocatalytic activity of the samples was evaluated by detecting changes in the intensity of the characteristic absorption bands of toxic impurities over time using the IR-spectroscopy technique (IR-spectrometer Bruker IFS-66v/S with the spectral range of 40–12000 cm$^{-1}$ and resolution of 0.5 cm$^{-1}$). A decrease in the intensity of the absorption bands of C-H bonds in toluene was measured under an incandescent lamp illumination.

### 3 Results and Discussion

According to the XRD data the samples under investigation were composed of the anatase phase. The morphology and crystal structure of N-doped and non-doped titania were studied using TEM. It was found that there were no variations in the shape and size of the nanoparticles after changing the dopant level during the synthesis. All the investigated samples consisted of irregular shaped crystals in the size range of 10–20 nm (Fig. 1).

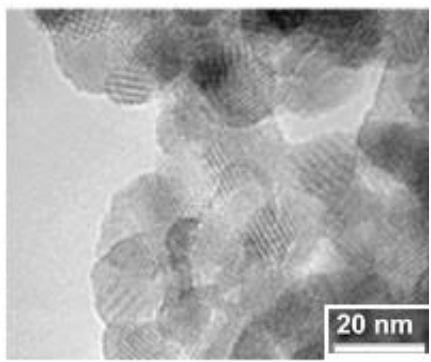

**Fig. 1** TEM image of N-TiO$_2$-3 sample

To investigate the chemical composition of the N-TiO$_2$ samples the Ti 2p, O 1s and N 1s core level spectra were measured by XPS, as shown in the figures below. The Ti 2p spectra were represented by one state: the binding energies of Ti 2p$_{3/2}$ peaks were close to each other for all the investigated N-doped samples and corresponded to the charge state of titanium in TiO$_2$ [33]. The O 1s spectra were also very similar for all the N-TiO$_2$ samples and were represented by two states with different intensities: (I) the peak with a lower binding energy, possibly corresponding to lattice oxygen in TiO$_2$, and (II) the peak with a greater energy, corresponding to OH-groups or chemisorbed oxygen on the surface [33]. Since there was no difference in the shape of the spectra indicated above, here we show them only for the N-TiO$_2$-1 sample (Fig. 2).

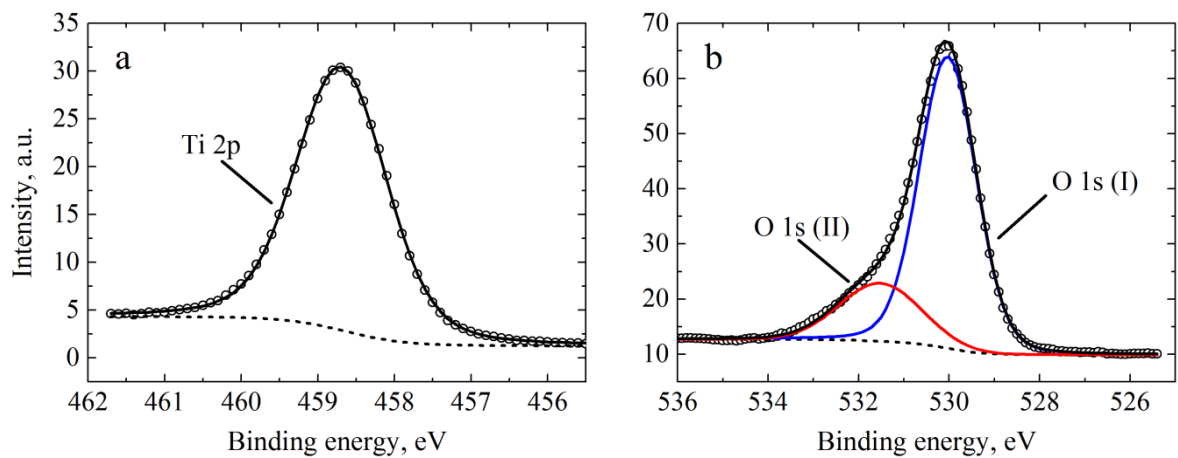

**Fig. 2** XPS spectra of Ti 2p (a) and O 1s (b) of N-TiO$_2$-1 sample

The N 1s spectra were represented by one state in the N-TiO$_2$-1 and N-TiO$_2$-2 samples, but in the N-TiO$_2$-3 sample a second state peak was found (Fig. 3). These peaks, according to the literature, correspond to interstitial (I) and O-substitutional (II) nitrogen in TiO$_2$ [33, 34].

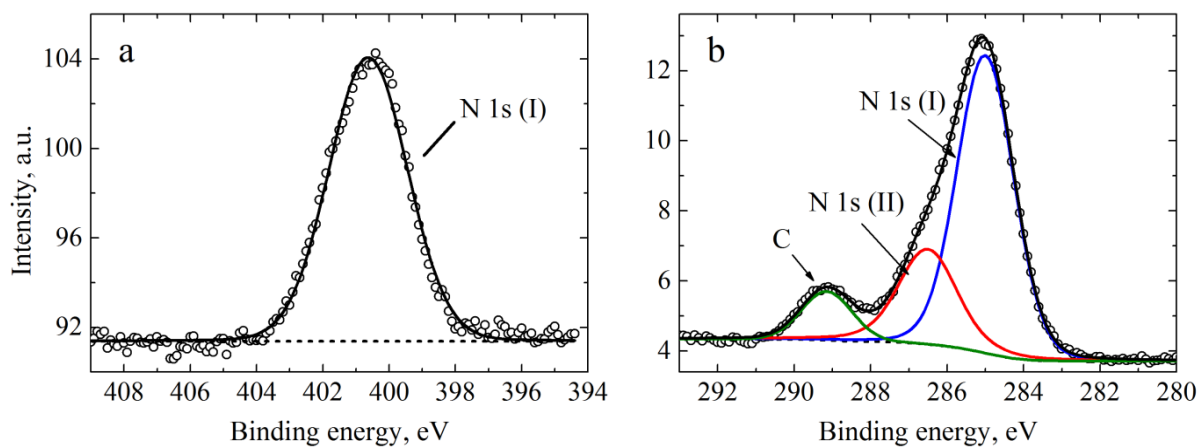

**Fig. 3** XPS spectra of N 1s of N-TiO$_2$-1 (a) and N-TiO$_2$-3 (b) samples

Table 1 summarizes the composition properties obtained from the XPS data of the N-TiO$_2$ samples.

To study the paramagnetic defects in N-TiO$_2$, the EPR spectroscopy measurements were carried out. It was found that the shape of the spectra, which is responsible for spin parameters of the defects, remained unchanged for all the doped samples, but their intensity increased with a rise in the nitrogen dopant level. The signal from the non-doped sample was faint, so we will not consider it.

**Table 1** Chemical composition of N-TiO$_2$ samples (sol-gel)

| Sample | Content, wt% | | |
|---|---|---|---|
| | Ti | O | N |
| N-TiO$_2$-1 | 27.2 | 57.7 (I)<br>14.9 (II) | 0.2 (I) |
| N-TiO$_2$-2 | 29.0 | 56.7 (I)<br>13.9 (II) | 0.4 (I) |
| N-TiO$_2$-3 | 27.8 | 55.9 (I)<br>15.3 (II) | 0.7 (I)<br>0.3 (II) |

The most intense spectrum obtained at 300 K is shown in Fig. 4 together with its computer simulation, calculated by means of EasySpin MATLAB toolbox. The spin-Hamiltonian parameters were as follows: the g-tensor — $g_1 = 2.0042$, $g_2 = 2.0031$, $g_3 = 2.0022$; the width of the EPR line — $\Delta H_1 = 3.7$ G, $\Delta H_2 = 2$ G, $\Delta H_3 = 2.8$ G; the hyperfine interaction constants — $A_1 = 2.3$ G, $A_2 = 3.3$ G, $A_3 = 33.0$ G. The EPR signal with such parameters, according to the literature data [33, 34], can be attributed to paramagnetic centers related to nitrogen atoms embedded into the titania matrix (nuclear spin I = 1) with an uncompensated electron spin. The maximal concentration of such N• radicals was $6\cdot10^{18}$ spin/g (in N-TiO$_2$-3 sample).

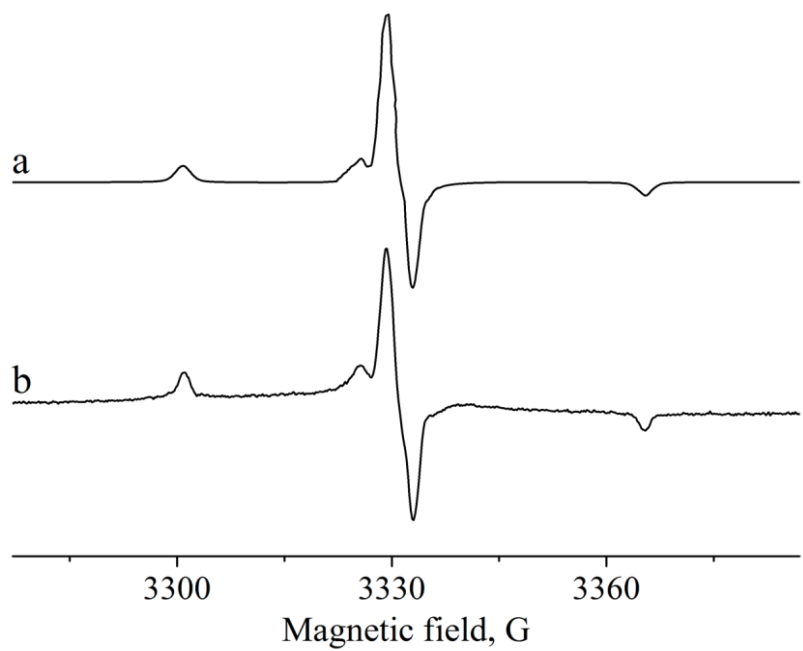

**Fig. 4** EPR spectrum of N-TiO$_2$-3 sample (b) and its computer simulation (a). T = 300 K

Despite the relatively high intensity, the signal described above was negligible, while the more intense signal appeared during the low-temperature (77 K) EPR experiments (Fig. 5). This strong signal was also simulated using the following parameters: the g-tensor — $g_1$ = 2.0002, $g_2$ = 1.99831, $g_3$ = 1.9280; the width of EPR line — $\Delta H_1$ = 4.4 G, $\Delta H_2$ = 3.9 G, $\Delta H_3$ = 22.0 G; the hyperfine interaction constants — $A_1$ = 0 G, $A_2$ = 33.5 G, $A_3$ = 4.0 G. Comparing this with the literature [34, 35] we can assume, that the observed signal corresponds to nitric oxide radicals (NO•), in which the unpaired electron is localized in a 2π-anti-bonding orbital.

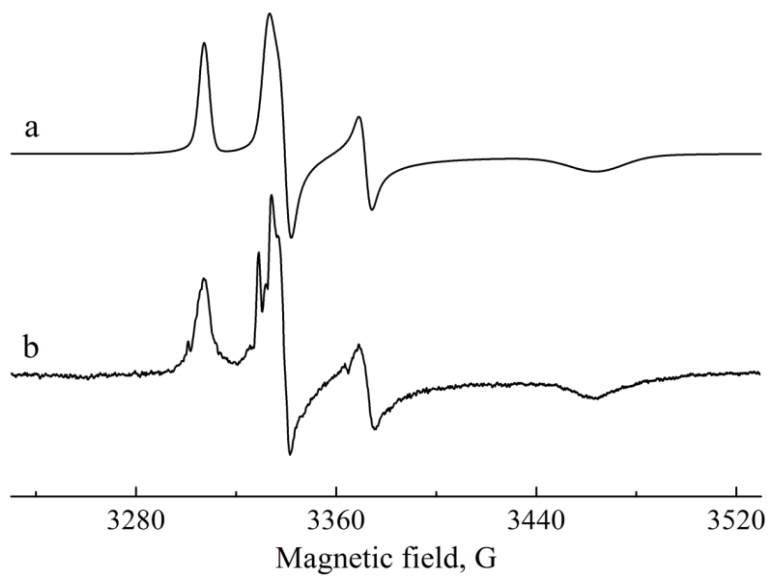

**Fig. 5** EPR spectrum of N-TiO$_2$-3 sample (b) and its computer simulation (a). T = 77 K

The temperature dependence of the shape of the EPR signal corresponding to NO• is shown in Fig. 6. The effect of temperature was completely reversible. The reason why NO• centers were not detected at 300 K lies in Heisenberg's uncertainty principle: the spin-lattice relaxation time $T_1$ of NO• radicals is very short, which leads to a huge uncertainty in energy, i.e. huge broadening of the EPR line. Decreasing the temperature one can increase the $T_1$ time to narrow the signal, which becomes visible and distinguishable [36]. The signal amplitude grows with the temperature drop according to Curie's law [36].

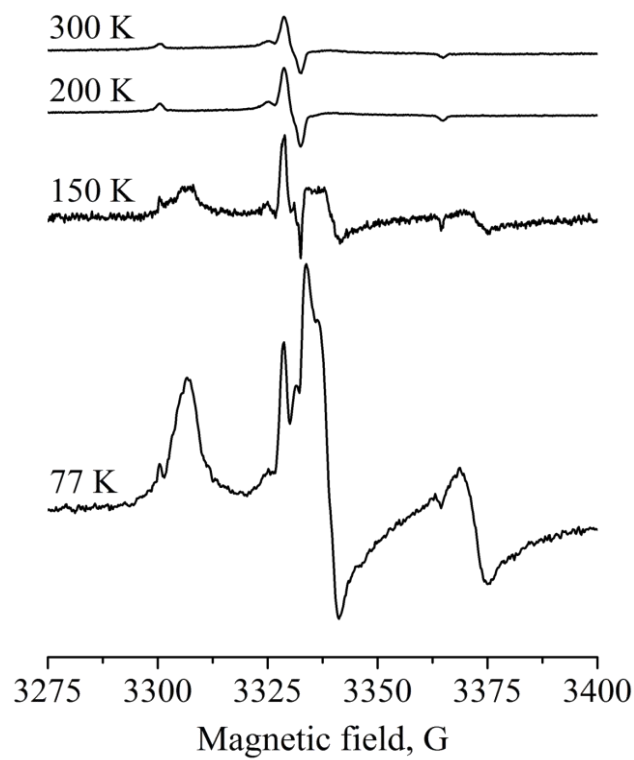

**Fig. 6** EPR spectra of N-TiO$_2$-3 sample measured at different temperatures

The maximal concentration of NO• radicals was $3\cdot10^{20}$ spin/g, which is almost two orders of magnitude higher than for N• radicals. The concentrations of both types of radicals were highly dependent on the amount of the dopant (Fig. 7).

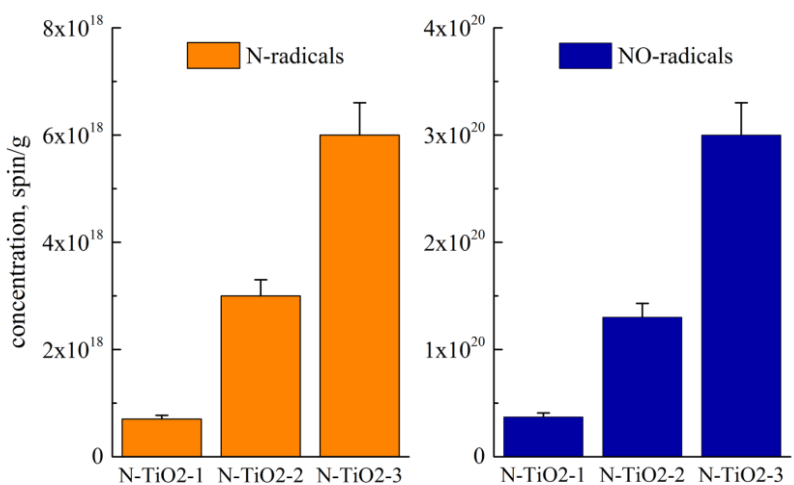

**Fig. 7** Concentration of N• and NO• radicals in N-TiO$_2$ samples

Illumination of all the samples with incandescent lamp led to an increase in the EPR signal intensity of N• radicals (Fig. 8a, the lowest curve) while the intensity of NO•-related signal slightly decreases (Fig. 8b, the lowest curve). The illumination effect was completely reversible, with the signal returning to its initial form in several minutes, which evidences light-induced recharge processes in the radical system.

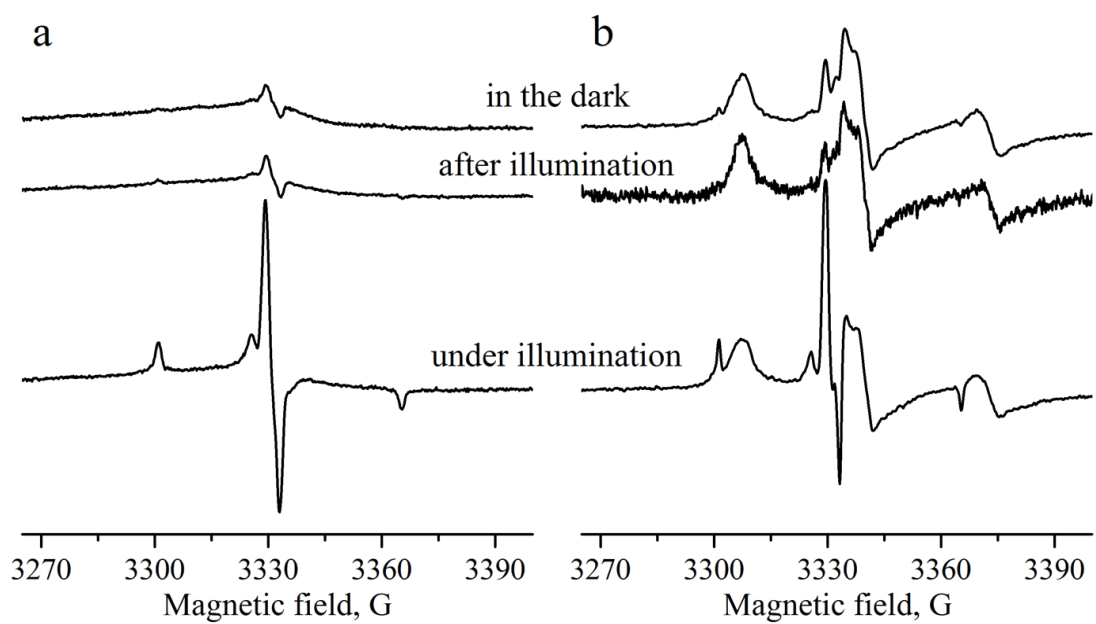

**Fig. 8** EPR spectra of N-TiO$_2$-3 sample in the dark and under illumination measured at 300 K (a) and 77 K (b)

The photoconductivity spectra were recorded for the direct detection of photoinduced charge carriers in the samples (Fig. 9). The non-doped sample has a typical peak with the maximum at ~3.15 eV ($\lambda \approx 400$ nm), which corresponds to the band gap of titania. But one can see that the curves from the N-doped samples have a tail in the range about 1.45–1.95 eV (850–400 nm).

The appearance of such tails indicates the formation of dopant (N-related) states in the band gap of N-doped titania. Thus, the impurity absorption occurs during the visible light irradiation of N-TiO$_2$. Electrons and holes photoexcited from the impurity levels contribute to the photocurrent, and the intensity of the described tails grows with the level of dopant. This can be explained by an increase in the number of nitrogen levels in the band gap of N-TiO$_2$, which leads to a rise in the concentration of charge carriers. Thus, the existence of doping-related defects in N-TiO$_2$ leads to the appearance of photoconductivity under the visible light illumination due to defect recharge processes.

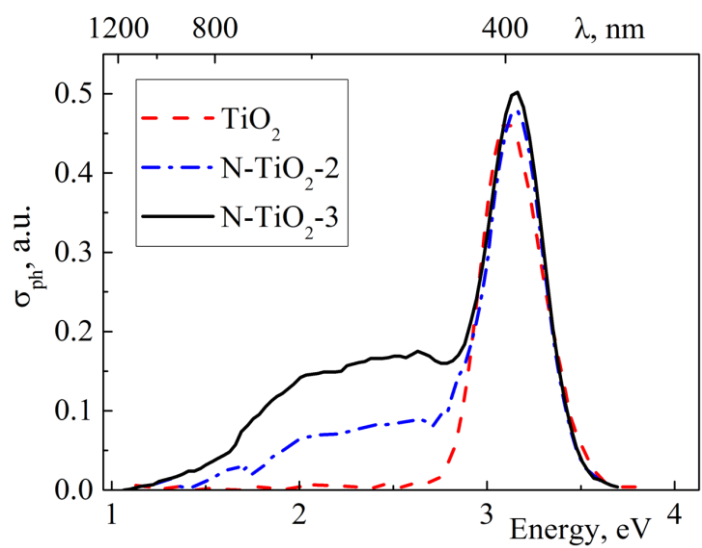

**Fig. 9** Photoconductivity spectra of non-doped and N-doped $TiO_2$ samples

Finally, the curves of the toluene photocatalytic decomposition under the visible light of an incandescent lamp were measured to identify the influence of defects not only on the electron transport properties of the samples, but also on the photocatalytic activity as an important practical property of titanium dioxide. The obtained results are presented in Fig. 10.

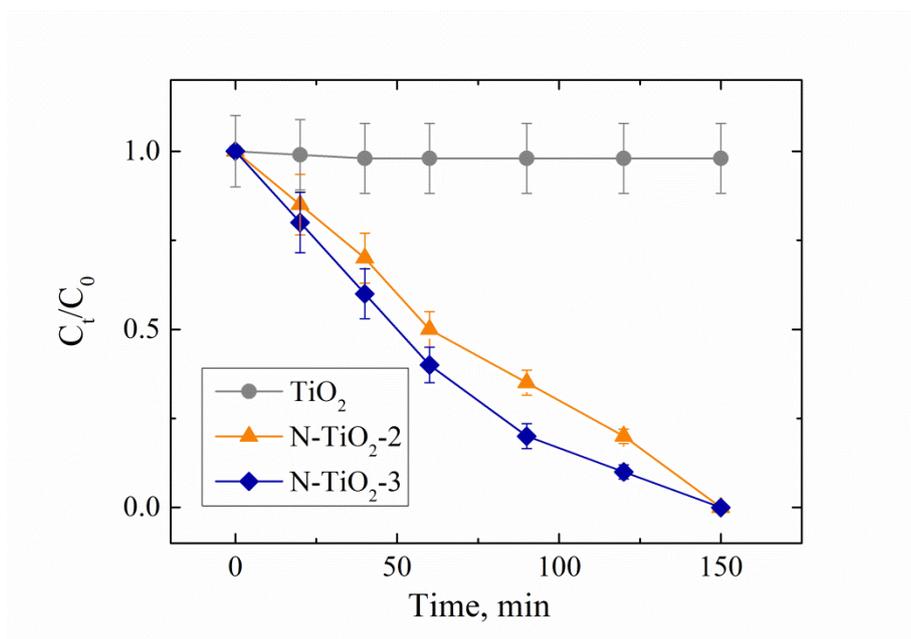

**Fig. 10** Photocatalysis kinetics of non-doped and N-doped $TiO_2$ samples under an incandescent lamp illumination in the presence of toluene molecules. $C_0$ and $C_t$ are the concentrations of the contaminant in the air at the time $t_0 = 0$ and $t$ respectively. $T = 300$ K

One can see that the experimental curve of the non-doped $TiO_2$ has an almost straight shape, which means a very weak photocatalytic activity of this sample under visible light. But the N-$TiO_2$ samples show a finite value of the decontamination time under the same conditions, the photocatalytic activity increasing with the growth of the dopant amount. This along with the photoconductivity data directly confirms that the paramagnetic defects found in the samples are responsible for the formation of energy levels in the titania band gap. We also suppose that the variations of the EPR signal intensity during illumination of the samples (Fig. 8) evidences that N-related radicals are involved in photocatalytic processes in N-$TiO_2$ but this question needs further investigation. The described results can be useful for the development of $TiO_2$-based photocatalytic applications.

## 4 Conclusions

Samples of nitrogen-doped titanium dioxide (anatase, $0.2 \leq N \leq 1.0$ wt%) prepared by the sol-gel method were investigated using X-band EPR spectroscopy, photoconductivity and photocatalysis experiments. N• and NO• radicals in N-$TiO_2$ have been observed at T = 300 K and 77 K respectively, their concentrations were calculated. The spin-Hamiltonian parameters of both types of defects were obtained from EPR spectra computer simulations. An increase in the generation rate of charge carriers was found in the N-doped samples by means of the photoconductivity measurements under visible light. These samples also had a good photocatalytic activity under visible light unlike the non-doped $TiO_2$. Therefore for the first time we have showed the presence of a correlation between photoconductivity, photocatalysis data and radical concentrations. The obtained results can be useful in further studies and photocatalytic applications.

**Acknowledgments**

This work was supported in part by the Russian Foundation for Basic Research (project no. 16-32-00800 mol_a) and by the Fund (Federal) for Assistance to Small Innovative Enterprises ("U.M.N.I.K." grant no. 5816GU2/2015 (0010844)). The measurements have been done using the facilities of the Collective Use Center at Lomonosov Moscow State University.